# Electronic Origin of Contrasting Li and Na Clustering on Graphene

Gabriel Kuderowicz[b] and Oleksandr I. Malyi*[a,b]

[a]*Qingyuan Innovation Laboratory, 1 Xueyuan Road, Quanzhou 362801, P. R. China*

[b]*Centre of Excellence ENSEMBLE3 Sp. z o. o., Wolczynska Str. 133, 01-919, Warsaw, Poland*

*Contact author: oleksandrmalyi@gmail.com

## Abstract

The electrochemical performance of anode materials is often defined by the ability of metal ions (e.g., Li and Na) to be effectively stored within the host material over a specific range of chemical potentials. This is particularly true for adsorbed Li and Na atoms on carbon surfaces (e.g., graphite or pore surfaces in hard carbon), where the interplay between ionic and metallic bonding often takes place - while metal atoms act as ionic species, in a finite concentration limit, they tend to form metallic bonding when they approach each other. What makes the case special is that, despite similar chemical identities, Li and Na atoms exhibit distinct tendencies for cluster formation; for example, the formation of Li metallic clusters on a graphene surface is only thermodynamically favorable after reaching a finite metal-ion concentration. Motivated by this, herein, we use first-principles calculations on graphene to examine the electronic origin of Li and Na clustering. Specifically, we demonstrate that the separated adsorbed atoms donate electrons to graphene, shifting the Fermi level to higher energies; however, the formation of metal clusters can result in the formation of lower-energy orbitals that are specifically associated with metal-metal bonding, thus lowering the Fermi level (and hence total energies). Importantly, while the process is somewhat similar for Na and Li atoms, the energy levels of the newly formed metallic states govern the difference in metallic cluster formation between Li and Na.

## Introduction

In metal-ion batteries, it is generally believed that metal ions are stored in an ionized state by donating their valence electrons to the host material during the formation of ionic bonds. This ionic picture provides a useful framework for understanding the battery operation voltage window, insertion sites, ion transport, configurational ordering (e.g., driven by the tendency to minimize ion-ion interactions due to strong electrostatic repulsion), and the concentration-dependent chemical potential. [1–3] However, it breaks down when the metal concentration is high, since the species that have been adsorbed can approach each other and form metallic clusters, which, while can induce extra capacity, can also serve as sites for dendrite formation.[4–6] This is especially important in relation to carbon electrodes (i.e., graphite, hard carbon) used for Li- and Na-ion storage, since it is well established that the intercalation of metal ions between the graphite layers generally leads to an ionic mechanism,[7–10] while in hard carbon, it is the formation of metal clusters that plays a critical role in metal-ion storage in the plateau-like regions.[6,11–15] The ionic picture described above can capture the physics of the dilute storage regime but fails to explain the clustering tendencies observed at higher concentrations. It is important to note that the transition from ion-like adsorption to metal-like aggregation/clustering is not simply the replacement of an ionic bond with a metallic one; rather, there is an intrinsic interplay between charge transfer and metal-metal clustering.

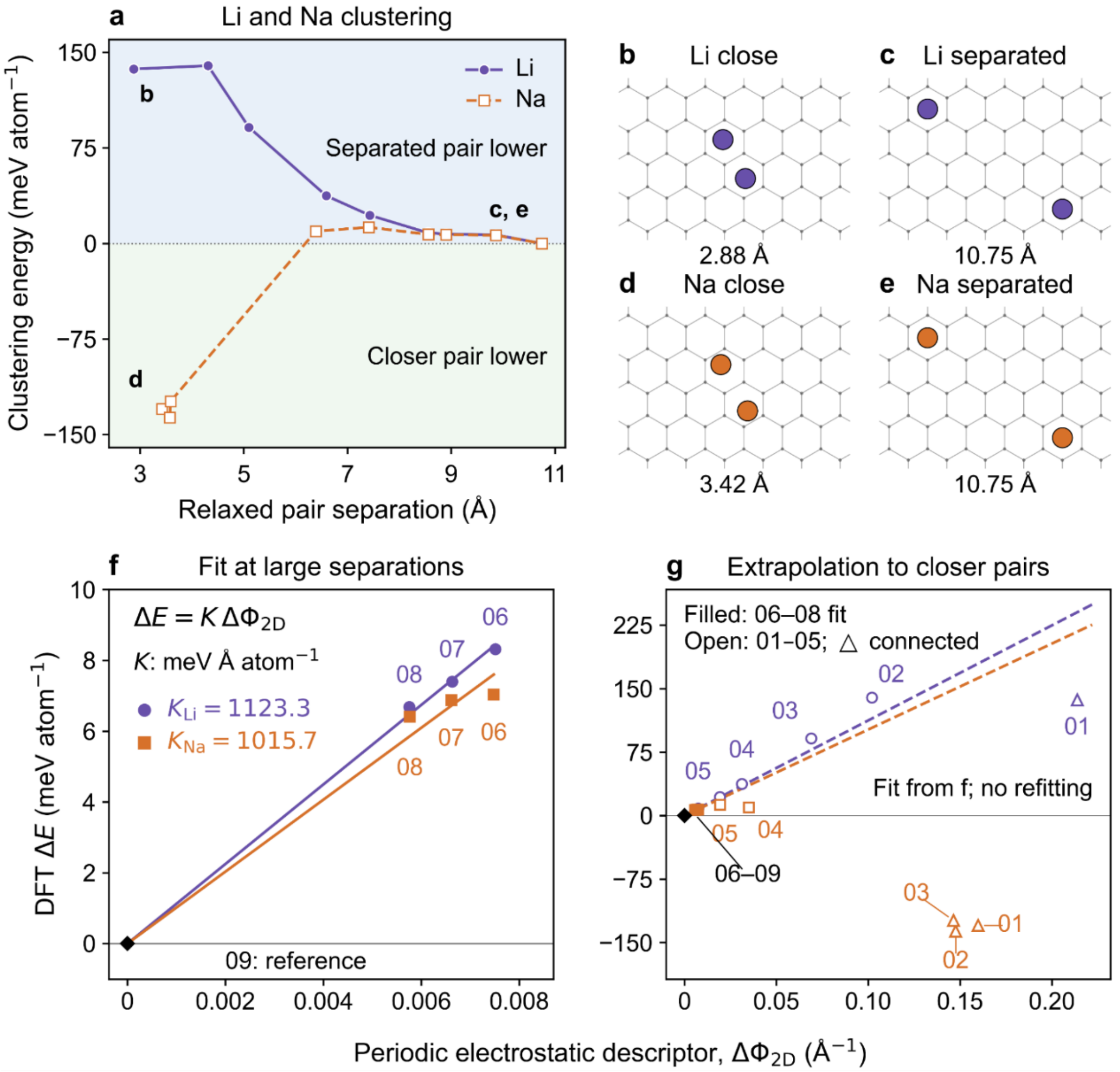

**Fig. 1** Clustering tendency between 2 Li/Na atoms on the graphene surface: (a) Clustering energy per adsorbed atom relative to the most separated configuration as a function of the distance between metal atoms. (b-e) Representative configurations for Li and Na on graphene surfaces. (f) Fit of DFT relative energies by relative Ewald summation for large separation and (g) its performance for other configurations. The numbering of atomic configurations is taken from the energy-to-distance plot (a), starting from the shortest separation.

Graphene provides a simple, well-defined carbon surface for examining this interplay without the additional effects of interlayer insertion or pore confinement and hence has been intensively used by different groups for exploring clustering tendencies,[16–19] where indeed it was demonstrated that both Li and Na tend to cluster at some critical concentration of metal ions, but show somewhat different chemistry. These studies attributed the difference directly to the interplay of metallic vs ionic bonding, without a deep understanding of the mechanism behind bond formation or what specifically causes the difference in clustering tendencies between Li and Na atoms. For instance, Figure 1 shows that, despite their similarity, two Na atoms already prefer to cluster, while, in contrast, Li tends to maximize the distance until the critical Li concentration is reached. This raises two fundamental questions: how do initially ionized, repelling metal ions develop the attractive interactions required for cluster formation, and why does this transition occur differently for Li and Na? Motivated by this, herein, we reexamine the origin of Li and Na clustering on pristine graphene using first-principles methods. We compare dispersed and clustered Lin and Nan configurations (n = 1–6) to examine their contrasting aggregation energies. For each number of adsorbed atoms, we consider dispersed, partially connected, and compact configurations and investigate their relaxed structures, relative energies, metal-metal connectivity, adsorption heights, and magnetic states. Specifically, we demonstrate that while well-separated metals indeed donate electrons to the graphene, raising the Fermi level, the formation of metallic clusters induces a metallic band that shifts high-energy occupied states to lower energies, thereby stabilizing the metallic cluster. While the chemistry of interactions between different metal ions and graphene is similar, it is the difference in electronic structure and interatomic interactions that defines the distinct clustering behaviors of Na and Li on graphene.

## Methodology

DFT calculations with spin polarization were performed using VASP,[20–22] with PAW pseudopotentials.[23,24] Van der Waals interactions were included in the optB88-vdW exchange-correlation functional.[25] Importantly, in many compounds (including graphite) it provides better agreement with experiments and more complex calculations[26,27] for quantities such as interlayer spacing and binding energy compared to the PBE-D2,[28] and PBE-D3 approximations,[29,30] which are often used. Semi-core states were treated explicitly, so there are 3 and 7 valence electrons in calculations for Li and Na, respectively. A plane-wave cutoff of 500 eV was chosen for both Li and Na.

Graphene supercells were considered with 25 Å of vacuum along the z-axis and an 8×8×1 size (128 C atoms). Self-consistent cycle calculations were performed on a 3×3×1 Γ-centered k-grid with a Gaussian smearing width equal to 0.05 eV, whereas the density of states (DOS) was calculated on a 9×9×1 grid. All calculations were converged to $10^{-7}$ eV for the total energy, and the geometry-optimization force threshold was 0.01 eV/Å. Adsorbed particles on graphene can induce a significant dipole moment, which results in much slower energy convergence with respect to the vacuum height in the supercell. Using dipole corrections[31,32] to address the issue has the additional advantage of obtaining the average local potential in vacuum: $V_{vac} = V_{ionic}(r) + \int dr'\rho(r')/|r - r'|$, where $V_{ionic}$ (r) is the ionic potential mimicked by pseudopotentials and ρ(r') is the electron density. By referencing the electronic structure to $V_{vac}$, the shifts in energy levels upon clustering can be quantitatively compared for different structures.

We tested all possible planar configurations for n=2-3 starting from $C_6$ ring positions and reported stacked structures. For higher concentrations, we analyzed only selected cluster geometries presented in Figs. S1-S3. However, forces in some Na structures were not converged up to 0.01 eV/Å, so those configurations were not included in the analysis.

In order to highlight the differences in the electrostatic repulsion of adsorbed Li and Na ion pairs on graphene we calculated the electrostatic energy of a periodic lattice of charges using the Ewald summation[33] implemented in Pymatgen.[34] For simplicity we assumed charges +1 on metal and 0 on C atoms for each relaxed structure in Fig 1a because we are discussing electrostatic repulsion in graphene's plane and the transferred charge accumulated on the surface is responsible for the perpendicular force pulling ions to graphene. Then with those geometries we calculated Ewald energies. The periodic electrostatic descriptor $\Delta\Phi_{2D}$ was obtained from the Ewald energy, referenced to the most separated structure, and divided by the Coulomb constant (charges on Li/Na are 1e): $\Delta\Phi_{2D}=(E^{Ewald}{}_{i}-E^{Ewald}{}_{far})/k_e$, $k_e$=14.3996 eV Å $e^{-2}$. It was used as an effective 1/r inverse distance to plot relative total energies of corresponding structures in Fig. 1 f-g.

## Results and discussion

Similar to the traditional picture, Na and Li atoms act as ionic species on the graphene surface, adsorbing at the center of the 6-membered C ring (i.e., hollow site). This adsorption results in charge donation from Li and Na to the surrounding atoms (Fig. S4), which is in line with well-established Na/Li-C interaction chemistry,[17–19] suggesting that the contrast between Li and Na therefore emerges when neighboring adsorbates interact, rather than from a change in the direction of charge transfer. Indeed, as the concentration of metallic species increases, the centers of the 6-membered C rings become inequivalent - metal atoms demonstrate different interaction chemistry depending on the interaction between them: when the atoms are far apart, similar to that for single atoms, Na acts mainly as an ionized species, and relative energies are mainly defined by ionic repulsion and the corresponding Na-Na distance. However, at smaller separations, Na configurations can relax into connected structures with Na-Na distances ranging from 3.42 to 3.59 Å and energies that are 123.9 to 136.6 meV per Na lower than those of the separated reference. This behavior is somewhat different from that of Li atoms, which, similar to Na, act as ionic species at larger separations, but at a distance of 2.88 Å, the Li configuration is still 136.8 meV per Li above the fully separated configuration. It is important to note that while the closest Li cluster configuration (2.88 Å) is less stable than the separated configuration, it is 2.6 meV per Li lower than the next one at 4.31 Å, suggesting that the difference between these two configurations cannot be described from the ionic perspective only. This is indeed demonstrated in Fig. 1f,g, which shows that the relative energy at short separations deviates significantly from the periodic Ewald model fitted at large distances. Specifically, at large separations, interaction energy follows the relative Ewald fit (i.e., $\Delta E = K\Delta\Phi_{2D}$, where K is a fitted proportionality coefficient and $\Delta\Phi_{2D}$ represents the change in the two-dimensional periodic Coulomb descriptor relative to the same-species 09 reference), but its fit performance worsens noticeably when close-atom configurations are accounted for, indicating that the electrostatic picture of interaction cannot describe interactions between metal atoms in close proximity, implying a change in the nature of interatomic interaction as the distance between atoms changes.

Considering the electronic structure of graphene, it is evident that the interaction of a single Li/Na atom with graphene shifts the Fermi level from the Dirac point to higher-energy states - metal ions act as n-type dopants, introducing free electrons to the conduction band (Fig. S4). Similarly, when two atoms are introduced and well separated, their main metal-associated states remain unoccupied above the Fermi level (Fig. 2d,e). This behavior resembles the traditional n-type doping picture,[35–37] in which both Li and Na atoms act as electron donors and higher dopant concentrations shift the Fermi level to higher energies. Nevertheless, as is known from doping physics, the introduction of free carriers is usually accommodated by compensation defects that can arise to pin the Fermi level.[38,39] While traditionally such compensation is usually done via formation of intrinsic defects - for instance, donor defects are compensated by intrinsic acceptors - we find that in the case of metal atoms, they themselves have the compensation mechanism - when metal atoms approach each other, the electron density reorganizes in a manner consistent with metal-metal bond formation. Figure 2a,b shows the calculated charge density difference ($\Delta\rho_{pair} = \rho(G+M_A+M_B) - \rho(G+M_A) - \rho(G+M_B) + \rho(G)$, where G denotes graphene, $M_A$ and $M_B$ are the two Li or Na adsorbates, and ρ is the electron density of each indicated system, with all atoms fixed at their positions in the paired structure), demonstrating electron accumulation between the Li or Na atoms accompanied by electron depletion in the region between the metal pair and graphene. This accumulation between the atoms reflects an electronic interaction beyond independent charge donation to graphene and typical ionic repulsion.

Importantly, this is activated only when the metal ions approach each other, thereby providing a real-space picture of the developing bond that can be related to the changes in metal-associated electronic states discussed below. Clustering of metal atoms also directly affects the C-Li/Na interaction, which is reflected in changes in corresponding distances. Thus, Li and Na distances to the graphene plane increase from 1.76 and 1.95 Å to 2.25 and 2.48 Å upon clustering, respectively, which directly implies a weakening of the corresponding interaction.

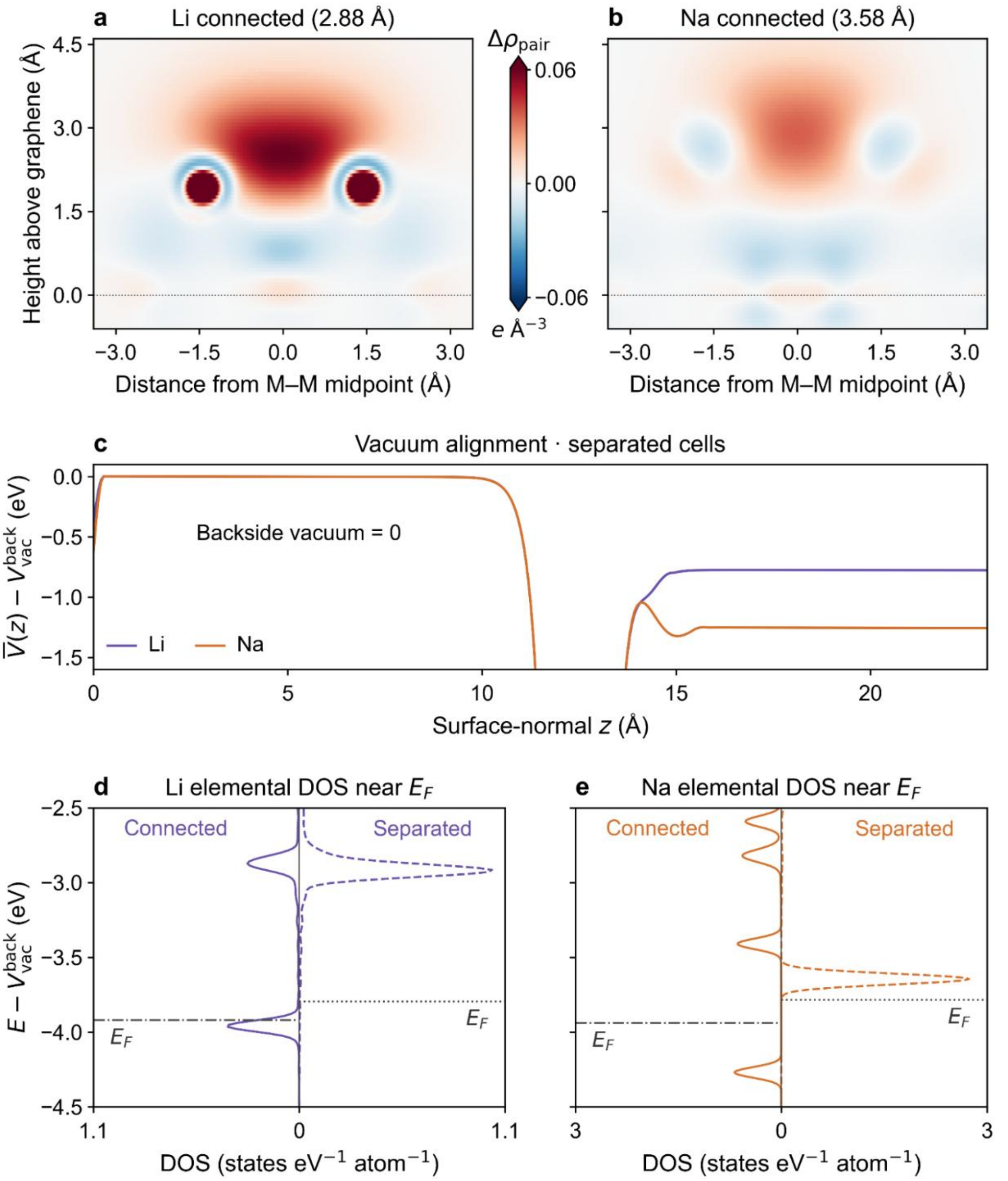

**Fig. 2** Electronic structure of 2-atom clusters: (a-b) Charge-density difference $\Delta\rho_{pair} = \rho(G+M_A+M_B) - \rho(G+M_A) - \rho(G+M_B) + \rho(G)$ for two metal ions interacting on the graphene surface, in plane with M atoms. (c) x-y planar average of the electrostatic potential for the separated Li and Na configurations, referenced to the vacuum on the opposite side of graphene as an illustration of the common reference state used for analysis of projected (d) Li and (e) Na densities of states in cluster and well-separated configurations.

The key electronic change upon formation of Li/Na clusters is the development of occupied metal-associated states that are not present when metal ions act separately. Specifically, for well-separated Li and Na atoms, the main metal-associated states lie 0.88 and 0.14 eV above the Fermi level, respectively, and remain unoccupied. In the cluster configurations, occupied metal-associated states develop below the Fermi level, consistent with the emergence of metal-metal bonding. However, referencing each configuration to its own Fermi level can actually hide changes in the electronic-state energies themselves due to differences in the interaction chemistry. Because of this, we therefore align the states to a common vacuum reference to distinguish their energy shifts from changes in the Fermi level itself. Taking into account that adsorption on the graphene surface actually changes the surface properties, for the alignment itself we select the surface side that does not contain metal atoms, as it is less affected by the presence of the adsorption of metal ions (Fig. 2c). The resulting aligned electronic structures are shown in Fig. 2d,e and show distinctly that the main Li-associated states in the connected state appear at -3.97 eV, while the corresponding main Na states are at -4.27 eV. The formation of these clusters also associates with the change in Fermi level energies as compared to the isolated energy (i.e., the intrinsic compensation mechanism discussed above), which is slightly larger for Na (0.15 eV) vs Li (0.12 eV) and consistent with redistribution of electrons to occupied metal bands. It is important to note that the appearance of occupied metal-associated states (i.e., Li and Na metallic states below the Fermi level) does not necessarily imply favorable clustering, as illustrated by the fact that two Li atoms energetically prefer a separated configuration, whereas Na atoms thermodynamically favor pairing. Nevertheless, the lower energy of the occupied Na-associated state correlates with the more favorable clustering tendency of Na, consistent with the proposed electronic origin of this difference. To further explore the identified difference in the electronic-structure origin of clustering, we examine clustering for a larger number of metal atoms.

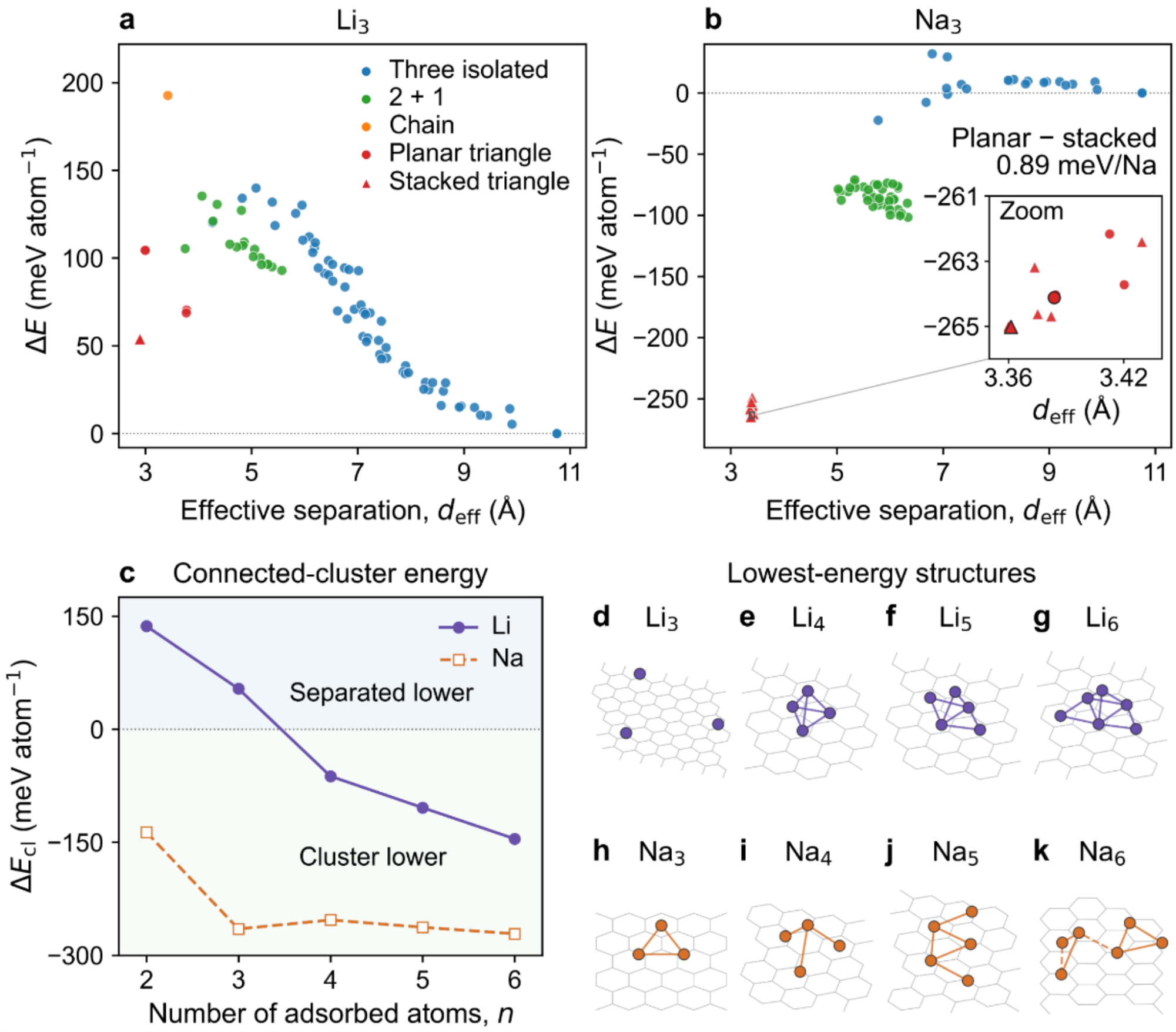

**Fig. 3** Size-dependent clustering of Li and Na atoms on graphene: (a-b) Relative energies of 3-atom Li and Na configurations as a function of the distance between metal atoms. The zoomed-in inset shows the corresponding sampled planar and stacked $Na_3$ configurations. (c) Clustering energy for different Li/Na cluster sizes and the identified lowest-energy (d-g) $Li_3$-$Li_6$ and (h-k) $Na_3$-$Na_6$ configurations.

As the number of adsorbed atoms increases, the competition between metal-graphene and metal-metal interactions allows a wider range of arrangements. This is already apparent for 3 atoms, where the adatoms can remain separated, form a dimer with an isolated atom (i.e., 2+1 configuration), or assemble into a connected cluster (i.e., approximately parallel to graphene or in a stacked arrangement extending away from the surface with corresponding labels given in Fig. 3a,b). All these configurations are unique in their structural arrangement and have different energies. Taking this into account, for simplicity in the following discussion, we consider relative energies (with respect to the most separated metal-atom configuration) as a function of the effective separation $d_{eff}(N) = N(N-1)/(2\sum_{i=1}^{N-1}\sum_{j=i+1}^{N} 1/d_{ij})$, where $d_{ij}$ is the distance between them, accounting for periodic boundary conditions) between N adsorbed atoms. This effective separation thus provides a common descriptor for measuring the proximity of the adsorbed atoms, while still allowing for variations in metal-metal and C-metal spatial arrangements, each of which can yield different interaction energies. The corresponding results in Figure 3 demonstrate that Na atoms tend to interact strongly with each other, forming a distinct planar cluster configuration with a clustering energy of -264 meV/Na, a substantially stronger tendency than that observed for 2 Na atoms (i.e., clustering energy of -137 meV/Na). The computed results here also demonstrate stair-like energy behavior: isolated atoms have the highest energies, partial clustering reduces the energy, followed by a range of closely spaced energy configurations. Thus, formation of a dimer with a single isolated Na atom lowers the energy by 102 meV/Na, while the formation of a planar triangular cluster produces a further reduction to -264 meV/Na. The lowest-energy sampled stacked $Na_3$ structure lies only about 1 meV per Na below (more favorable) the planar one; hence, it should be noted that the relatively weak interaction between metal atoms and graphene allows stabilization of range-similar clusters with somewhat comparable energies. For Li, the triangular form of the metal cluster also has lower energy than the dimer plus the single isolated Li atom (i.e., 2+1 configuration), but it remains higher in energy than the configuration containing three separated Li atoms. Among the compact $Li_3$ configurations, the lowest-energy sampled stacked arrangement is 15.2 meV per Li below the planar one. In this way, the additional connectivity increases the stability of the local cluster, but does not make its formation energetically favorable. The comparison of relative clustering energetics for 2 and 3 Li atoms suggests that increasing Li concentration reduces (makes more favorable) the clustering energy from +136.8 meV/Li to +53.7 meV/Li. This trend is further enhanced as the Li concentration increases further - the clustering energy changes from +53.7 meV/Li for the connected $Li_3$ configuration to -62.4 meV/Li for Li4, then to -104.1 and -145.4 meV/Li for $Li_5$ and $Li_6$, respectively. For Na, clustering remains energetically favorable as the number of adsorbed atoms increases (Fig. 3c), with clustering energies close to those of $Na_3$ configurations - the clustering energies are -253, -263, and -272 meV for $Na_4$, $Na_5$, and $Na_6$, respectively. These results thus demonstrate a difference in size dependence for Li and Na clustering on graphene surfaces. Specifically, Li clustering becomes progressively more energetically favorable over the investigated concentration range, reaching the crossing points between 3 and 4 metal atoms (it should be noted, however, that the above electronic structure picture of the model suggests that). For Na, much of the per-atom stabilization (relative to well-separated atoms) is already achieved in the connected 3-atom cluster, with comparatively little variation as the number of atoms increases. The electronic results for 2-atom clusters provide a microscopic picture of the interactions underlying these trends. Both metals develop occupied metal-associated states upon pairing, accompanied by redistribution of charge at the metal-graphene interface. The lower energy of the occupied Na-associated state is consistent with its favorable pairing energy, whereas the Li results show that the development of occupied metal-associated states can precede an energetic preference for clustering. The overall preference therefore depends on the balance between the development of metal-metal bonding and the accompanying changes in interactions with graphene.

## Conclusion

In summary, using first-principles theory, we systematically studied Li and Na clustering tendencies on the graphene surface, showing that this process is complex and involves competition between metallic and ionic band formation. We show that well-separated metal ions donate their electron to the conduction band, shifting up the Fermi level - larger concentration of metal atoms results in more electrons donated to the conduction band. As the metal-ion concentration increases, further increases in the Fermi level are pinned by metallic-cluster formation, resulting in metal-associated states and a downward shift of the Fermi level. Referencing the electronic structures to a common vacuum level shows that the occupied Na-associated state lies lower in energy than the corresponding Li-associated state, consistent with Na more favorable clustering tendency and providing an electronic-structure origin for the behavior. Specifically, while Na clustering is favorable already for 2 Na atoms, Li clustering requires a higher concentration. The Li results therefore demonstrate that occupied metal-associated states can develop before clustering becomes energetically favorable. The different clustering tendencies of Li and Na reflect the overall balance between the developing metal-metal bonding and the accompanying changes in interaction with graphene.

## Acknowledgements

Research on Na storage mechanisms in carbon-based materials was supported by the National Centre for Research and Development (NCBR) under project WPC3/2022/50/KEYTECH/2024. The Startup Fund of Qingyuan Innovation Laboratory (Grant No. 00524009) and the Fujian Province High-Level Innovative and Entrepreneurial Talents Program (Grant No. 30225003) partially supported the work in China. Computational resources and technical support were provided by the Polish high-performance computing infrastructure PLGrid through ACK Cyfronet AGH and WCSS under computational Grant No. PLG/2025/018614. Access to the LUMI supercomputer, owned by the EuroHPC Joint Undertaking and hosted by CSC (Finland) and the LUMI consortium, was provided through PLGrid under Grant No.

PLL/2025/09/018974. Institutional support for ENSEMBLE3 was provided through the “Centre of ExcelleNce for nanophotonicS, advancEd Materials and novel crystal growth-Based technoLogiEs – ENSEMBLE3” project, carried out within the 2.1 International Research Agendas programme of the Foundation for Polish Science and co-financed by the European Union under the European Funds for Smart Economy 2021–2027 (FENG), and through the European Union Horizon 2020 research and innovation programme, Teaming for Excellence (Grant Agreement No. 857543).